# Specifics of the Elemental Excitations in "True-1D" MoI₃ van der Waals Nanowires

Fariborz Kargar[1,*], Zahra Barani[1], Nicholas R. Sesing[2], Thuc T. Mai[3], Topojit Debnath[4], Huairuo Zhang[5,6], Yuhang Liu[4], Yanbing Zhu[7], Subhajit Ghosh[1], Adam J. Biacchi[8], Felipe H. da Jornada[9], Ludwig Bartels[10], Tehseen Adel[3], Angela R. Hight Walker[3], Albert V. Davydov[6], Tina T. Salguero[2], Roger K. Lake[4], and Alexander A. Balandin[1,*]

[1]Nano-Device Laboratory (NDL) and Phonon Optimized Engineered Materials (POEM) Center, Department of Electrical and Computer Engineering, University of California, Riverside, California 92521 USA

[2]Department of Chemistry, University of Georgia, Athens, Georgia 30602 USA

[3]Quantum Measurement Division, Physical Measurement Laboratory, National Institute of Standards and Technology (NIST), Gaithersburg, Maryland 20899 USA

[4]Laboratory for Terascale and Terahertz Electronics (LATTE), Department of Electrical and Computer Engineering, University of California, Riverside, California 92521 USA

[5] Theiss Research, Inc., La Jolla, California 92037, USA

[6]Materials Science and Engineering Division, National Institute of Standards and Technology, Gaithersburg, Maryland 20899 USA

[7]Department of Applied Physics, Stanford University, Stanford, California 94305 USA

[8]Nanoscale Device Characterization Division, Physical Measurement Laboratory, National Institute of Standards and Technology (NIST), Gaithersburg, Maryland 20899 USA

[9]Department of Materials Science and Engineering, Stanford University, Stanford, California 94305 USA

[10]Department of Chemistry, University of California, Riverside, California 92521 USA

* Corresponding authors: fkargar@ece.ucr.edu (F.K.); balandin@ece.ucr.edu (A.A.B); web-site: http://balandingroup.ucr.edu/

## Abstract

We report on the temperature evolution of the polarization-dependent Raman spectrum of exfoliated MoI$_3$, a van der Waals material with a "true one-dimensional" crystal structure that can be exfoliated to individual atomic chains. The temperature evolution of several Raman features reveals anomalous behavior suggesting a phase transition of a magnetic origin. Theoretical considerations indicate that MoI$_3$ is an easy-plane antiferromagnet with alternating spins along the dimerized chains and with inter-chain helical spin ordering. The calculated frequencies of the phonons and magnons are consistent with the interpretation of the experimental Raman data. The obtained results shed light on the specifics of the phononic and magnonic states in MoI$_3$ and provide a strong motivation for future study of this unique material with potential for spintronic device applications.

Recently a new research field focused on one-dimensional (1D) van der Waals (vdW) quantum materials has emerged from the earlier work on low-dimensional systems.[1–4] The 1D vdW materials are based on 1D structural motifs and include transition metal trichalcogenides and halides.[1–5] It is helpful to distinguish *true*-1D versus *quasi*-1D vdW systems. We define the material to be true-1D if it contains covalent bonds only in the direction of the atomic chains, with all other bonds being of vdW type; the material is quasi-1D if it contains strong covalent bonds along the 1D chain direction, while also bonded by weaker covalent bonds in the perpendicular planes. For example, according to this criterion, $Nb_2Se_9$ is a true-1D material whereas $TaSe_3$ is a quasi-1D material.[6–8] The number of vdW materials with quasi-1D or true-1D structures is not known, but machine learning investigations indicate that hundreds of 1D vdW materials are synthetically accessible.[5,9–11] True-1D vdW materials can be exfoliated chemically or mechanically into individual atomic chains or a few-atomic chain bundles.[2,6,12] One can also envision that progress in chemical vapor deposition (CVD) will eventually allow the controlled synthesis of such 1D materials on substrates of choice, thereby enabling their practical applications.[13–15]

Owning to their unique dimensionality and electronic density of states, 1D materials often support unusual quasiparticle excitations, such as separated spin-charge excitations, Luttinger plasmons with quantized velocities,[16] magnetism,[17] charge density waves, and emergent topology[18–21]. Many of such unusual electronic properties also manifest themselves as nontrivial features in the dispersion relation of other collective excitations, such as phonons and magnons, due to the lower spatial dimensionality and stronger electronic correlation in 1D. The magnetic properties of 1D materials are of particular interest. The intrinsic ferromagnetic (FM) and antiferromagnetic (AFM) spin orderings have been demonstrated in the single atomic layers of some transition-metal dichalcogenides ($MX_2$) and transition-metal phospho-trichalcogenides ($MPX_3$) that constitute quasi-two-dimensional (2D) systems.[22,23] In these compositions, "M" is a transition metal, "X" is a chalcogen, and "P" is phosphorous. The possibility of AFM ordering in true-1D vdW materials is interesting from both fundamental science and practical applications points of view. The prospect of the helical spin order, *i.e.*, spin-spiral order, in 1D materials is particularly exciting.[24] The challenges with the experimental observation of magnetism in the 1D limit are complex owing to the scarcity of stable true-1D vdW material systems, low signal-to-noise ratio in magnetometry and other related experiments, and difficulties in handling individual atomic chains.

As a step in this direction, we investigated elemental excitations in exfoliated van der Waals nanowires of $MoI_3$, classified as a true-1D vdW material, using Raman spectroscopy.[12] The latter can be useful for spintronic applications; one can envision the atomic chains of the $MoI_3$ serving as ultimately downscaled magnonic interconnects conducting spin currents.[25]

$MoI_3$ crystals for this study were prepared from the elements using the chemical vapor transport (CVT) technique.[12,26] The inclusion of $NH_4I$ as a transport agent led to a more reliable synthetic procedure, owing to lower internal ampule pressures, and excellent product quality, which allowed us to determine the single crystal structure of $MoI_3$ by x-ray diffraction; details and atomic coordinates are provided in the Supplemental Materials. Figure 1 (a) presents the crystal structure of $MoI_3$ from different views. Although this structure can be refined in both *Pmmn* and *P6₃/mcm* with low *R* values, we ultimately used *Pmmn* in line with the interpretation of powder x-ray diffraction (PXRD) data by Meyer and coworkers.[26] The P*mmn* structure was further confirmed by our high-angle annular dark-field scanning transmission electron microscopy (HAADF-STEM) imaging. The choice of space group is critical because *Pmmn* leads to "dimerized" $MoI_3$ chains showing clear Peierls distortion with alternating Mo–Mo distances of 2.8793(19) and 3.5298(19) Å, whereas *P6₃/mcm* leads to 1D chains of equidistant Mo atoms. We also validated the space group by PXRD; Figure 1 (b) shows the experimental pattern from $MoI_3$ crystals with selected *hkl* indices (see Supplementary Information for the complete *hkl* assignment). Refined lattice parameters from the single crystal study [$a$ = 12.3183(4) Å, $b$ = 6.4091(2) Å, $c$ = 7.1180(3) Å] are similar to those from our PXRD analysis [$a$ = 12.3200(6) Å, $b$ = 6.4088(6) Å, $c$ = 7.1225(3) Å] as well as literature values.[26] Figure 1 (c, d) presents the low-magnification HAADF-STEM and atomic-resolved image of the exfoliated $MoI_3$ crystals. The atomic-resolved model of $MoI_3$ confirms the *Pmmn* lattice structure along the [100] projection. Note that the scanning electron microscopy (SEM) image of representative $MoI_3$ crystals, presented in Figure 1 (e), shows that $MoI_3$ crystals grow preferentially along the *b*-axis, leading to needlelike structures with high aspect ratios. The energy dispersive spectroscopy (EDS) mapping of the exfoliated samples demonstrates homogeneous distribution and overlap of Mo and I (Figure 1 (f, g)). The EDS analytical results also confirm the expected composition of ~1:3 Mo:I (see Supplementary Information).

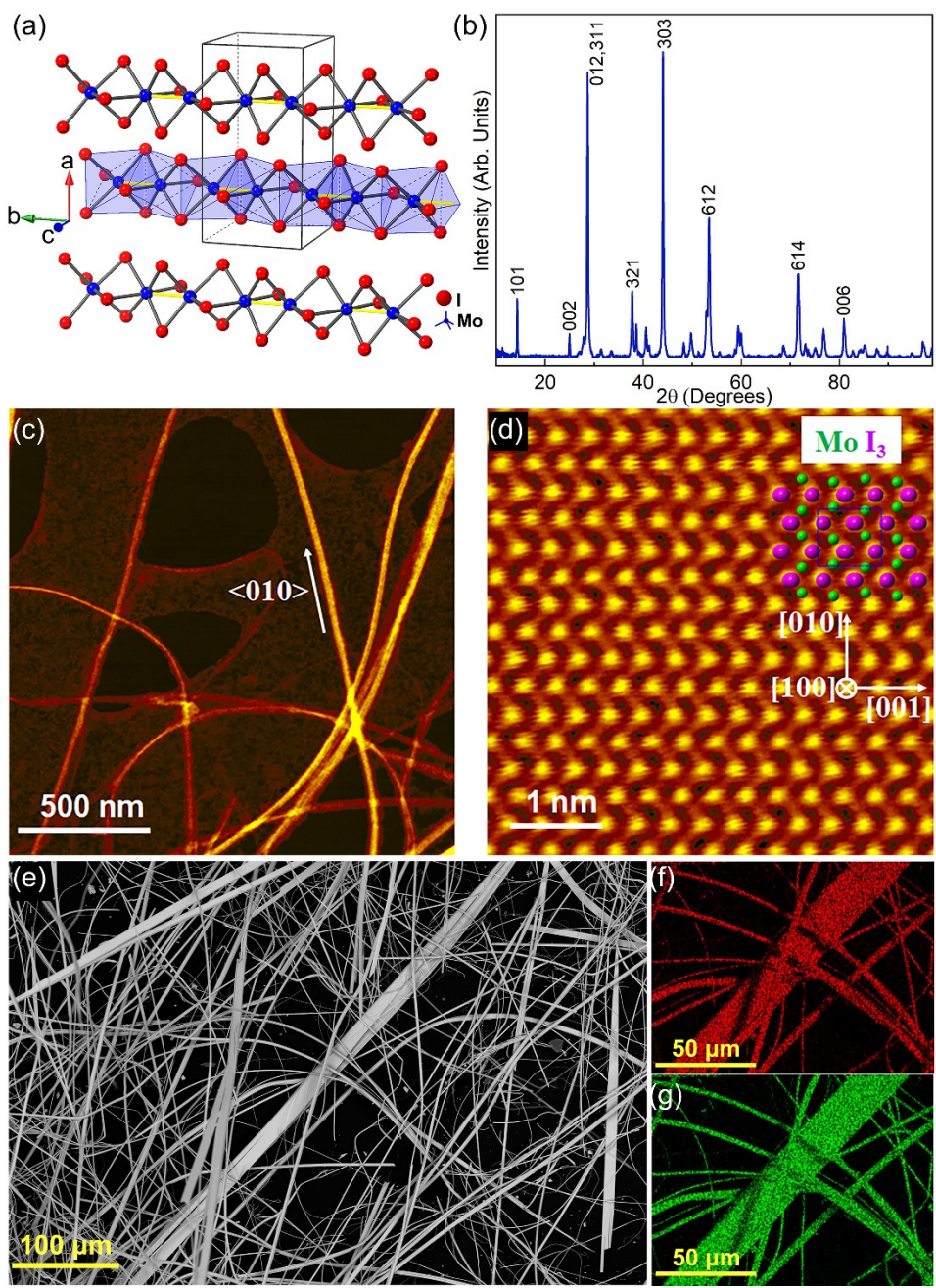

[Figure 1: (a) Crystal structure of MoI₃ refined in the P*mmn* (No. 59) space group; blue and red spheres represent Mo and I, respectively. The short Mo–Mo interactions of the "dimerized" MoI₃ chains are indicated with yellow bonds. (b) Experimental PXRD $\theta - 2\theta$ scan with selected *hkl* indices. (c) Low magnification HAADF-STEM image and, (d) atomic resolved HAADF-STEM image of exfoliated MoI₃ nanowires. The images confirm that the nanowires grow along the <010> direction. (e) SEM image of the as-synthesized MoI₃. (f, g) Corresponding EDS mapping of the MoI₃ crystal confirming the uniform distribution of Mo (red) and I (green).]

We used temperature and polarization-dependent Raman spectroscopy to study the elemental excitations, *i.e.,* phonons and magnons, in $MoI_3$ crystals. The experiments were conducted in the conventional backscattering configuration using a 488 nm laser excitation. In all experiments, the samples were positioned in such a way that the polarization of the incident light was along the "*b*" crystallographic direction, *i.e.*, along the nanowires. The polarization of the scattered light was selected to be either parallel with or transverse to the incident light. The power of the laser was kept at 400 µW to avoid self-heating effects. The results of the Raman measurements for the parallel and cross polarization arrangements in the temperature range of 1.6 K to 300 K are presented in Figure 2 (a) and (b), respectively. The unit cell of $MoI_3$ has 16 atoms with 45 optical phonon branches, resulting in a rich Raman response with many peaks in the range of 10 $cm^{-1}$ to 750 $cm^{-1}$. These Raman features strongly depend on the respective polarization of the incident and scattered light which is typical for one-dimensional materials.[27,28] As theoretically shown later in this manuscript, the maximum energy of optical phonons and magnons in $MoI_3$ at the Brillouin zone (BZ) center is ~228 $cm^{-1}$ and ~310 $cm^{-1}$, respectively. Therefore, the peaks observed in the spectra above ~310 $cm^{-1}$ are associated with the higher order Raman scattering processes, with the exception of the Si substrate peak at ~520 $cm^{-1}$. We plotted the spectral position and full-width-at-half-maximum (FWHM) of several well-defined peaks in the low-frequency and high-frequency regions, labeled by "$p_1$", "$p_2$" and "$p_3$" (Figure 2 (a)), as a function of temperature. These peaks have A$g$, B$_1$g, and B$_3$g vibrational symmetries, respectively. The results are presented in Figure 2 (c, d).

Generally, the phonon features in Raman spectra of different materials shift monotonically to lower frequencies with increasing temperature, which is explained by the crystal lattice expansion and phonon anharmonic effects resulting in softening of the phonon modes.[29] This is the case for the Raman peak at ~210 $cm^{-1}$. The spectral position of the peaks in the low-frequency region, at ~57 $cm^{-1}$ and ~88 $cm^{-1}$, blueshifts with increasing temperature. Phonon hardening with increasing temperature, observed for the peaks labeled as $p_1$ and $p_2$, is unusual. Similar anomalous behavior has been previously reported for the so-called soft phonon modes in ferroelectric materials such as $SrTiO_3$, and the transverse acoustic modes in Invar alloys and magnetic random solid solutions such as Fe$_{1-x}$Pd$_x$.[30,31] In the latter case, the phonon hardening was attributed to thermal magnetic fluctuations.[32] This observation suggests that MoI$_3$ nanowires reveal magnetic phenomena. One can further notice a deviation of the FWHM temperature dependence from a linear behavior, with a pronounced kink at ~125 K for both Raman features (p$_1$ and p$_2$). Such non-monotonic characteristic of FWHM likely indicates a phase transition in MoI$_3$.

The intensity of the peaks associated with scattering processes involving magnons tends to weaken with temperature and eventually fall to zero beyond the phase transition temperature. One can see in Figure 2 (a, b) that the intensity decreases for all peaks with increasing temperature, leaving the task of distinguishing possible magnon peaks from phonon peaks extremely difficult. It should be noted that the spectral position of the peaks is insensitive to the magnetic field of 9 T applied along crystal's "c" axis as well. Yet, this does not exclude the possibility of magnetic order. It is known that in many AFM materials, *e.g.*, NiO, even a strong magnetic field produce small effects on Raman spectrum.[33] Three broad Raman features at ~465 cm$^{-1}$, ~580 cm$^{-1}$, and ~625 cm$^{-1}$ (labeled as m$_1$, m$_2$, and m$_3$) encompassed by the rectangular dashed line in Figure 2 (a), contain additional valuable information. Theory described below indicates that these spectral features are associated with the two-magnon scattering processes. They disappear completely in the temperature range from 125 K to 150 K, and their energy exactly matches the calculated magnon energy at the BZ edge along the $\Gamma - Y$ direction, multiplied by a factor of two, as it should be for the two-magnon processes.

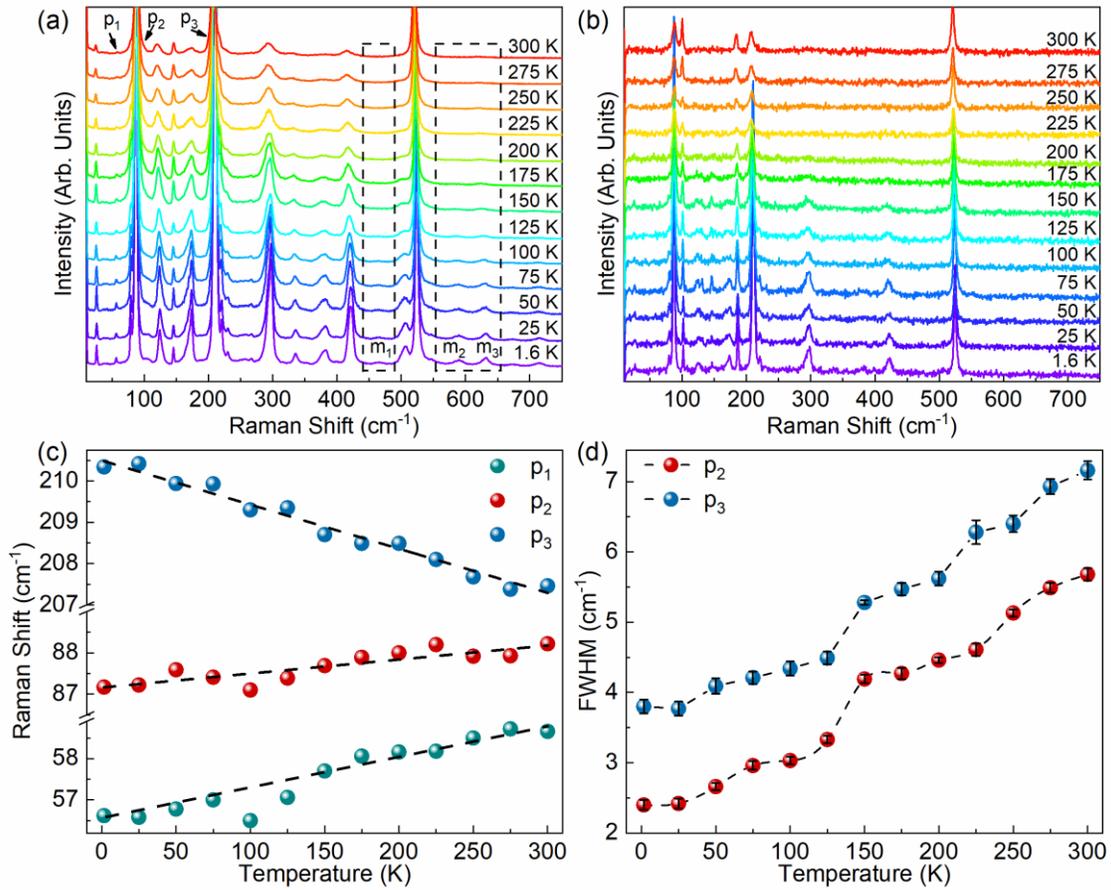

[Figure 2: Evolution of the (a) parallel, and (b) cross-polarization Raman spectra excited with 488 nm excitation as a function of temperature. The polarization of the incident light is along the nanowires. The dashed rectangular shows the high-frequency feature assigned to the two-magnon scattering process. (c) Spectral position of the peaks labelled as $p_1$, $p_2$, and $p_3$ in panel (b). (d) FWHM of the intense $p_2$ and $p_3$ Raman peaks at ~88 cm⁻¹ and 207 cm⁻¹ as a function of temperature. Note the abrupt increase in the FWHM of both peaks at 150 K.]

To further elucidate the nature of the observed phase transition in MoI₃, we plot the normalized intensity color-map of the Raman spectra shown in Figure 2 (a) as a function of temperature and frequency, in the low and high wavenumber regions. The results are presented in Figure 3 (a, b). The difference in the intensity of the Raman peaks in these two frequency ranges is significant; explaining the reason for plotting them separately. In both plots, the color change from purple to red indicates the peaks are becoming more intense. The white regions in Figure 3 (a, b) are the strong peaks at 88 cm⁻¹ and 207 cm⁻¹ and Si peak at 520 cm⁻¹ (see Figure 2 (a)). They are shown with white color because their intensity is beyond the defined maximum level in the colormap (the red color) to avoid overshadowing the weak Raman features of interest. One can see in Figure 3 (a) that none of the peaks has its intensity vanishing as the temperature rises. This confirms that all these peaks are associated with the phonon scattering processes. Interestingly, the broad Raman features at 465 cm$^{-1}$, 580 cm$^{-1}$ and 625 cm$^{-1}$ completely vanishes at the same temperature range. Similar broad features associated with the two-magnon scattering processes are common in AFM materials such as MnF$_2$, NiO$_2$, MnPSe$_3$, and VI$_3$.[34–37]

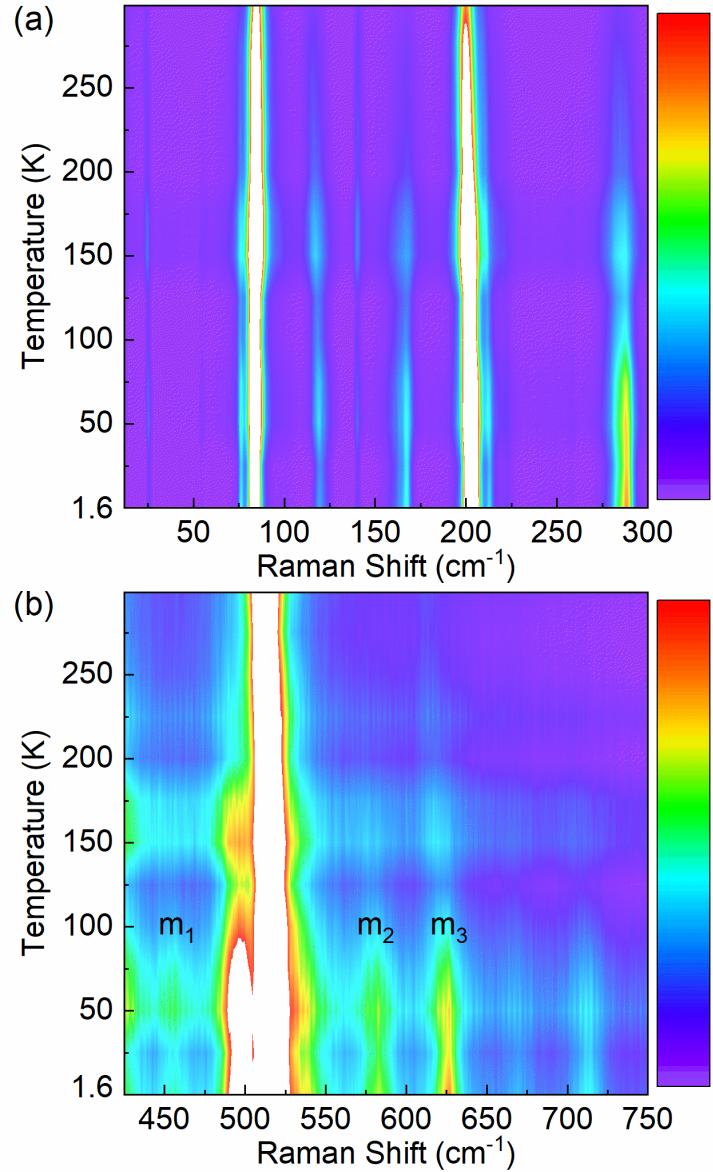

[Figure 3: The normalized intensity colormap of the Raman spectra accumulated in the parallel-polarization configuration. The spectra are presented for the (a) low and (b) high wavenumber ranges for clarity. Note that the intensity of the $m_1$, $m_2$, and $m_3$ peaks attributed to the two-magnon scattering gradually disappears at T~200 K.]

A recent study predicted the AFM spin-ordering in MoI₃ using machine learning algorithms.[5] This report did not describe the specific type of the AFM ordering or the Neel temperature of the transition in MoI₃. To compare directly the theory prediction with experiment, we performed our own *ab initio* calculations of the phonon and magnon dispersion of MoI₃. Density functional theory (DFT) calculations were carried out with the generalized gradient approximation (GGA)[38] exchange-correlation functional of Perdew, Burke, and Ernzerhof (PBE)[39,40] as implemented in the Vienna Ab initio Simulation Package (VASP).[41] The ion–electron interactions use the projected augmented wave (PAW) method. The energy cutoff for the plane wave basis was set to 520 eV. All the structures were allowed to relax during this process with a conjugate gradient algorithm until the energy on the atoms was less than $10^{-4}$ eV. The DFT-D3 approach proposed by Grimme[42] was used for the vdW interactions. A Hubbard +U value of 4.0 was used on the Mo atoms. This value was found to minimize the negative modes in the phonon dispersion. Phonon calculations were performed using the finite-displacement supercell approach as implemented in Phonopy.[43,44] For the phonon calculations, the Brillouin zone was sampled by a 4 × 7 × 4 Monkhorst–Pack k-point grid. The results of the calculations are presented in Figure 4 (a). The red spheres are the spectral position of Raman peaks in the range of 10 to 250 cm$^{-1}$ in the parallel polarization configuration (see Figure 2 (a)). The theoretical calculations are in excellent agreement with the experimental data.

The magnetic properties were calculated with PBE+U, as described above, and with the inclusion of spin-orbit coupling (SOC). To determine the magnetic ground state, total energy calculations were performed with the magnetic moments of the Mo atoms in ferromagnetic (FM) order and two different anti-ferromagnetic (AFM) orders. The ground state energy of the FM order was approximately 1.5 eV higher per unit cell than the ground state energies of the two different AFM configurations. In both AFM configurations, alternating spins were anti-aligned along the chains. From total energy calculations with different spin configurations, the exchange coupling constants and the magnetic anisotropy energy were determined for the magnetic Hamiltonian given by the following equation.

$$H_m = \sum_{\alpha m \neq \beta m} J_{\alpha m;\beta n} \, \boldsymbol{S}_{\alpha m} \cdot \boldsymbol{S}_{\beta n} + \sum_{\alpha m} A \left( S_{\alpha m}^y \right)^2 + g \mu_B \sum_{\alpha m} \boldsymbol{B} \cdot \boldsymbol{S}_{\alpha m}. \tag{1}$$

Here, Greek indices label the different chains, and roman indices label the atoms within the chain. There are 3 non-zero values for the exchange constants $J_{\alpha m;\beta n}$. Within the same chain, $J_1 = 187$ meV and $J_2 = 0.884$ meV are the exchange couplings corresponding to the shorter Mo-Mo distance and longer Mo-Mo distance, respectively, as illustrated in Figure 4 (b). $J_3 = 0.119$ meV is the exchange coupling between Mo atoms in the same plane on different chains, as illustrated in Figure 4 (b, c). All exchange terms are positive. The positive magnetic anisotropy energy (MAE) $A = 0.17$ meV, causes the spins to lie in the $a$-$c$ plane perpendicular to the chain direction. Magnetic anisotropy within the $a$-$c$ plane is too small to be resolved from our DFT calculations. The third term of the Hamiltonian is the external magnetic field. MoI₃ consists of dimerized AFM chains arranged in a triangular lattice as shown in Figure 4 (b-c). Total energy calculations based on the magnetic Hamiltonian of Eq. (1) found that the chain-to-chain spins in the same plane form spin spirals with 120° rotation angles between chains, typical of a triangular lattice. The magnon dispersion for bulk MoI₃ was calculated from Eq. (1) using the linear spin wave theory as implemented in SpinW[45] and is shown in Figure 4 (d). The numbers in this plot are the frequencies of different magnon branches at high symmetry points. The red spheres are the spectral position of the two-magnon Raman peaks shown in Figure 3 (b) divided by two. The frequency of the two-magnon peaks exactly matches the energy of the magnon branches at the BZ edge along the $\Gamma - Y$ direction. We can state that the theory of the phonon and magnon states in this true-1D vdW material is consistent with our interpretation of the experimental Raman data.

We have also theoretically considered the effects of external magnetic field on the magnon dispersion. By introducing in the theory a magnetic field of 10 T, we have seen that the bands at gamma points are split by a small value of ~0.5 cm⁻¹, which was not present in the dispersion without the field. The experimental efforts to detect this difference in the energies of the split bands were not conclusive since the frequncy differences were the limit of the instrumental resolution of our Raman systems. The Raman spectra collected under a magnetic field of 9 T at 1.6 K do not exhibit any pronounced changes in the spectral position of the peaks nor in splitting of any magnon peaks. This experimental observation is in line with the theory that predicted only small splitting of bands at the high magnetic field. Our preliminary magnetometry data revealed a weak signal. The latter can be attributed to difficulty of the measurements of the exfoliated nanowires. Further studies are needed with the larger quantities of the material for the direct comparison of the magnetometry and Raman data.

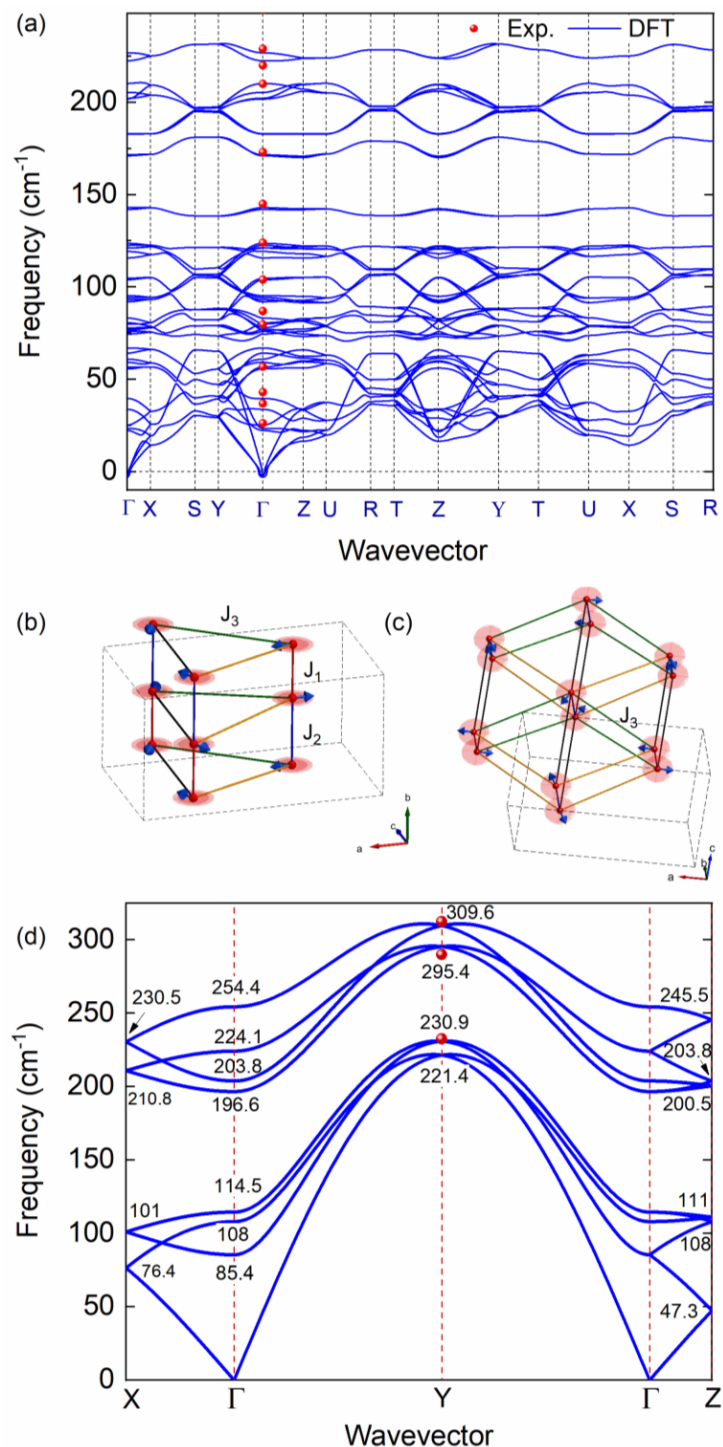

[Figure 4: (a) Calculated phonon dispersion with the experimental peaks shown by the red spheres. (b) Triangular arrangement of chains with 120° rotation of spins between nearest neighbor chains. (c) Dimerized chains illustrating the two different intra-chain exchange constants, $J_1$ and $J_2$, and the inter-chain exchange constant $J_3$. The red circles indicate the easy plane of the magnetic anisotropy. (d) Calculated magnon dispersion with the frequencies at the high symmetry points shown.]

In summary, we described here the specifics of the elemental phonon and magnon excitations in crystals of MoI$_3$, a vdW material with a *true*-1D crystal structure. Our measurements revealed anomalous temperature dependence of the Raman spectral features of MoI$_3$, which suggest magnetic phase transitions. The theoretical calculations of the phonon and magnon dispersion are in excellent agreement with the experimental data. The magnon zone-edge frequencies are consistent with the interpretation of the Raman data and the assignment of the two-magnon peaks. According to the theoretical considerations, the AFM spin texture has an easy plane perpendicular to the chains with alternating spins along the chains. The lowest energy arrangement for inter-chain ordering resulted from a spin spiral arrangement of nearest-neighbor inter-chain spins. The results shed light on the phononic and magnonic properties of 1D MoI$_3$ nanowires and provide a strong motivation for future study of this unique material with potential for spintronic device applications. One can envision that the nanowires or individual atomic chains of the electrically insulating MoI$_3$ becoming ultimately downscaled interconnects for magnonic spin currents.

## Acknowledgments

A.A.B. was supported by the Vannevar Bush Faculty Fellowship from the Office of Secretary of Defense (OSD) under the Office of Naval Research (ONR) contract N00014-21-1-2947 on One-Dimensional Quantum Materials. The work of T.T.S. and R.K.L. was supported by the subcontract of this ONR award. F.H.J., A.A.B. and L.B. also acknowledges the support from the National Science Foundation (NSF) program Designing Materials to Revolutionize and Engineer our Future (DMREF) *via* a project DMR-1921958 entitled Collaborative Research: Data-Driven Discovery of Synthesis Pathways and Distinguishing Electronic Phenomena of 1D van der Waals Bonded Solids. This work used the Extreme Science and Engineering Discovery Environment (XSEDE), which is supported by National Science Foundation Grant No. ACI-1548562 and allocation ID TG-DMR130081. H.Z. acknowledges support from the U.S. Department of Commerce, NIST under financial assistance award 70NANB19H138. A.V.D. acknowledges support from the Material Genome Initiative funding allocated to NIST.

## Disclaimer

Certain commercial equipment, instruments, software or materials are identified in this paper in order to specify the experimental procedure adequately. Such identifications are not intended to imply recommendation or endorsement by NIST, nor it is intended to imply that the materials or equipment identified are necessarily the best available for the purpose.

## Contributions

A.A.B. and F.K. conceived the idea, coordinated the project, contributed to experimental data analysis, and led the manuscript preparation. N.R.S. synthesized bulk crystals by CVT. T.T.S supervised material synthesis and contributed to data analysis. Z.B., T.T.M., and T.A. conducted Raman spectroscopy studies at UCR and NIST. H.Z. conducted the TEM characterization; A.R.H.W. and A.V.D. supervised material characterization and contributed to data analysis. T.D., Y.L., and Y.Z. conducted *ab initio* simulations of the phonon and magnon dispersion at UCR and Stanford, respectively. F.H.J. and R.L. contributed to the data analysis and supervised the simulations. L.B., S.G., and A.J.B. contributed to data analysis. All authors contributed to the manuscript preparation and editing. T.T.S. and A.V.D. thank Dr. Pingrong Wei (UGA) and Dr. Sergiy Krylyuk (NIST) for assistance with the single crystal and powder x-ray diffraction study.

## Data Availability Statement

The data that support the findings of this study are available from the corresponding author upon reasonable request.

## Supplemental Information

The supplemental information with additional data on material synthesis and characterization is available at the Applied Physics Letters journal web-site for free of charge.

## Conflict of Interest

The authors declare no conflict of interest.

# Supplemental Information

# Specifics of the Elemental Excitations in "True-1D" MoI₃ van der Waals Nanowires

Fariborz Kargar[1,*], Zahra Barani[1], Nicholas R. Sesing[2], Thuc T. Mai[3], Topojit Debnath[4], Huairuo Zhang[5,6], Yuhang Liu[4], Yanbing Zhu[7], Subhajit Ghosh[1], Adam J. Biacchi[8], Felipe H. da Jornada[9], Ludwig Bartels[10], Tehseen Adel[3], Angela R. Hight Walker[3], Albert V. Davydov[6], Tina T. Salguero[2], Roger K. Lake[4], and Alexander A. Balandin[1,*]

[1]Nano-Device Laboratory (NDL) and Phonon Optimized Engineered Materials (POEM) Center, Department of Electrical and Computer Engineering, University of California, Riverside, California 92521 USA

[2]Department of Chemistry, University of Georgia, Athens, Georgia 30602 USA

[3]Quantum Measurement Division, Physical Measurement Laboratory, National Institute of Standards and Technology (NIST), Gaithersburg, Maryland 20899 USA

[4]Laboratory for Terascale and Terahertz Electronics (LATTE), Department of Electrical and Computer Engineering, University of California, Riverside, California 92521 USA

[5] Theiss Research, Inc., La Jolla, California 92037, USA

[6]Materials Science and Engineering Division, National Institute of Standards and Technology, Gaithersburg, Maryland 20899 USA

[7]Department of Applied Physics, Stanford University, Stanford, California 94305 USA

[8]Nanoscale Device Characterization Division, Physical Measurement Laboratory, National Institute of Standards and Technology (NIST), Gaithersburg, Maryland 20899 USA

[9]Department of Materials Science and Engineering, Stanford University, Stanford, California 94305 USA

[10]Department of Chemistry, University of California, Riverside, California 92521 USA

---

* Corresponding authors: fkargar@ece.ucr.edu (F.K.); balandin@ece.ucr.edu (A.A.B); web-site: http://balandingroup.ucr.edu/

**Contents**

# 1 Chemical vapor transport synthesis and growth of MoI₃ crystals

Within an Ar-filled glovebox, 0.0470 g (0.324 mmol) of $NH_4I$ powder (Fisher Scientific, 99.0%) was placed at the bottom of a nitric acid-cleaned and dried ~18 x 1 cm fused quartz ampule (10 mm inner diameter, 14 mm outer diameter, volume of ~13 $cm^3$). This was followed by 0.8166 g (6.435 mmol) of $I_2$ crystals (JT Baker, 99.9%) and then by 0.2052 g (2.139 mmol) Mo powder (Strem, 99.95%). Clean transfer was assisted by a glass funnel and anti-static brush. After capping the ampule with an adapter and valve, the assembly was removed from the glovebox. While submerged in an acetonitrile/dry ice bath, the ampule was evacuated four times with Ar backfilling on a Schlenk line before being sealed under vacuum. The ampule was placed in a horizontal tube furnace and over 4 h, the temperature was ramped up to establish a gradient of 360 °C (source zone) – 300 °C (growth zone). After maintaining this gradient for 240 h, the ampule was cooled to room temperature over 6 h. 31.0 mg of lustrous silver, wire-like crystals were recovered from the growth zone (3.04% isolated yield). These crystals were subsequently stored within an Ar-filled glovebox.

# 2 Characterization details

Powder X-ray diffraction (PXRD) measurements were collected using a Bruker D8 Advance diffractometer equipped with an EIGER2 R 500 K single-photon-counting detector. Data were acquired in the Bragg–Brentano geometry from 10° to 100° 2θ using monochromatic Cu Kα (λ= 1.5418 A) radiation. The powder was prepared by grinding and mounting MoI$_3$ crystals on the zero background diffraction silicon plates. PXRD pattern was analyzed using MDI Jade version 6.5 software.

Powder X-ray diffraction (PXRD) data were collected on a Bruker D8 Advance instrument utilizing a Co−Kα X-ray source (λ = 1.78890 Å) operated at 35 kV and 40 mA. Data were collected from 10 to 80° 2θ with a scan rate of 0.2 s/step. Sampled crystals were prepared as pressed mounts, with wires being aligned in the same direction. Scanning electron microscopy (SEM) imaging was performed with an FEI Teneo FE-SEM at 10 keV with a spot size of 13. Energy dispersive x-ray spectroscopy (EDS) was performed using an Aztec Oxford Instruments X-MAX$^{N}$ detector operated at 10 keV with a spot size of 13. Crystals were mounted onto a stub with carbon tape and mechanically exfoliated using tape just prior to analysis.

**Table S1:** EDS results of as-synthesized MoI$_3$ crystals

|  | Mo (at. %) | I (at. %) |
|---|---|---|
| Theoretical | 25.0 | 75.0 |
| Experimental | 23.6 | 76.4 |

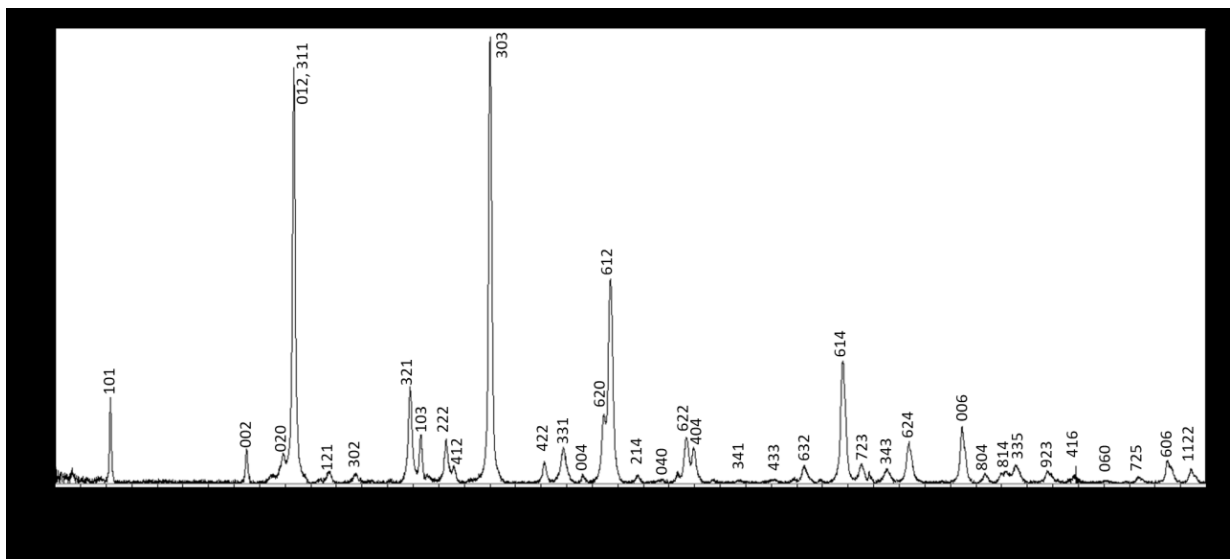

**Figure S1:** Experimental MoI₃ PXRD $\theta - 2\theta$ scan with *hkl* indices in the P*mmn* (No. 59) space group. The refined orthorhombic lattice parameters a=12.3200(6) Å, b=6.4088(6) Å, c=7.1225(3) Å.

# 3 Crystal structure of MoI₃

A specimen of MoI₃, approximate dimensions 0.030 mm × 0.030 mm × 0.300 mm as a dark silver crystal, was used for the single-crystal X-ray crystallographic analysis. The X-ray intensity data were measured at room temperature (295 K) on a Bruker D8 Quest PHOTON 100 CMOS X-ray diffractometer system with Incoatec Microfocus Source (IμS) monochromated Mo Kα radiation (λ = 0.71073 Å, sealed tube) using phi and omega-scan technique.

The integration of the data using an orthorhombic unit cell yielded a total of 21445 reflections to a maximum θ angle of 32.48° (0.66 Å resolution), of which 1134 were independent (average redundancy 18.911, completeness = 99.7%, $R_{int}$ = 4.38%, $R_{sig}$ = 2.35%) and 1055 (93.03%) were greater than $2\sigma(F^2)$. The final cell constants of $a$ = 12.3183(4) Å, $b$ = 6.4091(2) Å, $c$ = 7.1180(3) Å, volume = 561.96(3) Å$^3$, are based upon the refinement of the XYZ-centroids of 9937 reflections above 20 σ(I) with 6.62° < 2θ < 86.86°. The data were integrated with the manufacturer's SAINT software and corrected for absorption effects using the Multi-Scan method (SADABS). The calculated minimum and maximum transmission coefficients (based on crystal size) are 0.3171 and 0.7503.

The structure was solved and refined using the Bruker SHELXTL Software Package [S1], using the space group *Pmmn* (No. 59), with $Z = 4$ for the formula unit, MoI$_3$. Non-hydrogen atoms were located from successive difference Fourier map calculations. In the final cycles of each refinement, all the non-hydrogen atoms were refined in anisotropic displacement parameters. The final aniso-tropic full-matrix least-squares refinement on F$^2$ with 28 variables converged at R1 = 5.12%, for the observed data and wR2 = 13.38% for all data. The goodness-of-fit was 1.068. The largest peak in the final difference electron density synthesis was 3.404 e$^-$/Å$^3$ (0.70 Å from Mo(1)) and the largest hole was -2.161 e$^-$/Å$^3$ (1.67 Å from I(1)) with an RMS deviation of 0.401 e$^-$/Å$^3$. These largest residues are of no chemical significance. On the basis of the final model, the calculated density was 5.634 g/cm$^3$ and F(000), 804 e$^-$. The asymmetric unit contains a component in the form of Mo$_{0.50}$I$_{1.50}$ in a polymeric structure with all the atoms residing on symmetries (the site symmetry is: .m. for Mo(1), I(2), I(3) and mm2 for I(1), I(4)). The whole formula in the unit cell is in the form of MoI$_3$. The efforts have been made to resolve as many alerts as possible generated by CheckCIF. The current highest alerts are at level B, which are false alarms.

**Table S2:** Crystal data and structure refinement for MoI$_3$

| | |
|---|---|
| Empirical formula | MoI$_3$ |
| Formula weight | 476.64 |
| Temperature | 295(2) K |
| Wavelength | 0.71073 Å |
| Crystal system, space group | Orthorhombic, *Pmmn* |
| Unit cell dimensions | $a$ = 12.3183(4) Å  α = 90° |
| | $b$ = 6.4091(2) Å  β = 90° |
| | $c$ = 7.1180(3) Å  γ = 90° |
| Volume | 561.96(3) Å$^3$ |
| Z, Calculated density | 4,  5.634 Mg/m$^3$ |
| Absorption coefficient | 18.642 mm$^{-1}$ |
| F(000) | 804 |
| Theta range for data collection | 2.862 to 32.477° |
| Limiting indices | -18≤h≤18, -9≤k≤9, -10≤l≤10 |
| Reflections collected / unique | 21445 / 1134 [R(int) = 0.0438] |
| Completeness to theta = 25.242 | 99.5 % |
| Absorption correction | Semi-empirical from equivalents |
| Max. and min. transmission | 0.7503 and 0.3171 |
| Refinement method | Full-matrix least-squares on F$^2$ |
| Data / restraints / parameters | 1134 / 0 / 28 |
| Goodness-of-fit on F$^2$ | 1.068 |
| Final R indices [I>2σ(I)] | R1 = 0.0512, wR2 = 0.1314 |

| R indices (all data) | R1 = 0.0533, wR2 = 0.1338 |
|---|---|
| Extinction coefficient | 0.0026(5) |
| Largest diff. peak and hole | 3.404 and -2.161 e.A$^{-3}$ |

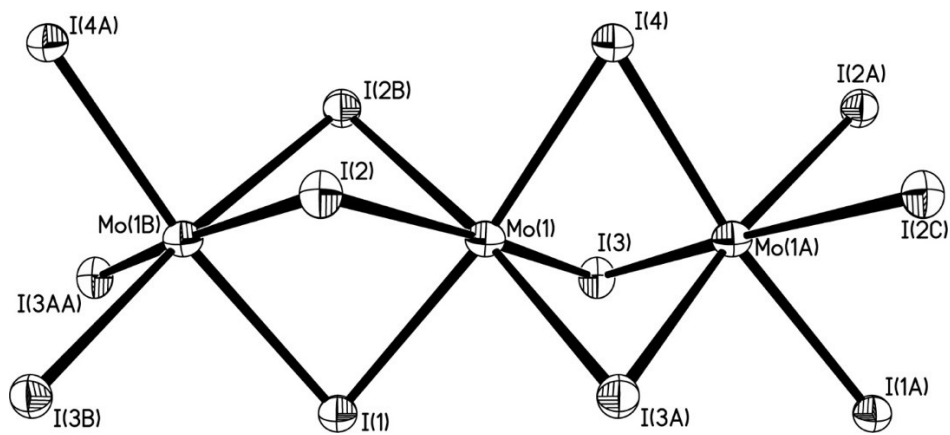

**Figure S2:** Crystal structure of MoI₃ (ORTEP representation)

**Table S3:** Atomic coordinates ($\times 10^4$) and equivalent isotropic displacement parameters ($A^2 \times 10^3$)

|  | x | y | z | U(eq) |
|---|---|---|---|---|
| Mo(1) | 7500 | 4746(2) | 7446(1) | 33(1) |
| I(1) | 7500 | 7500 | 10534(2) | 37(1) |
| I(2) | 9040(1) | 7500 | 5904(1) | 38(1) |
| I(3) | 5856(1) | 2500 | 9078(1) | 40(1) |
| I(4) | 7500 | 2500 | 4165(2) | 41(1) |

Note: U(eq) is defined as one third of the trace of the orthogonalized Uij tensor.

**Table S4:** Bond lengths [Å] and angles [°]

| Mo(1)-I(3)#1 | 2.7435(8) | Mo(1)-I(2)#2 | 2.8141(9) |
|---|---|---|---|
| Mo(1)-I(3) | 2.7435(8) | Mo(1)-I(1) | 2.8189(12) |
| Mo(1)-I(4) | 2.7434(12) | Mo(1)-Mo(1)#1 | 2.8793(19) |
| Mo(1)-I(2) | 2.8141(9) | Mo(1)-Mo(1)#2 | 3.5298(19) |
| | | | |
| I(3)#1-Mo(1)-I(3) | 95.18(4) | I(3)-Mo(1)-Mo(1)#1 | 58.35(2) |
| I(3)#1-Mo(1)-I(4) | 94.89(3) | I(4)-Mo(1)-Mo(1)#1 | 58.35(2) |
| I(3)-Mo(1)-I(4) | 94.89(3) | I(2)-Mo(1)-Mo(1)#1 | 128.840(19) |
| I(3)#1-Mo(1)-I(2) | 89.79(2) | I(2)#2-Mo(1)-Mo(1)#1 | 128.840(19) |
| I(3)-Mo(1)-I(2) | 172.81(4) | I(1)-Mo(1)-Mo(1)#1 | 128.76(2) |
| I(4)-Mo(1)-I(2) | 89.84(3) | I(3)#1-Mo(1)-Mo(1)#2 | 121.65(2) |
| I(3)#1-Mo(1)-I(2)#2 | 172.81(4) | I(3)-Mo(1)-Mo(1)#2 | 121.65(2) |
| I(3)-Mo(1)-I(2)#2 | 89.79(2) | I(4)-Mo(1)-Mo(1)#2 | 121.65(2) |
| I(4)-Mo(1)-I(2)#2 | 89.84(3) | I(2)-Mo(1)-Mo(1)#2 | 51.160(19) |
| I(2)-Mo(1)-I(2)#2 | 84.80(4) | I(2)#2-Mo(1)-Mo(1)#2 | 51.160(19) |
| I(3)#1-Mo(1)-I(1) | 89.90(3) | I(1)-Mo(1)-Mo(1)#2 | 51.24(2) |
| I(3)-Mo(1)-I(1) | 89.90(3) | Mo(1)#1-Mo(1)-Mo(1)#2 | 180.0 |
| I(4)-Mo(1)-I(1) | 172.89(4) | Mo(1)-I(1)-Mo(1)#2 | 77.52(5) |
| I(2)-Mo(1)-I(1) | 84.91(3) | Mo(1)-I(2)-Mo(1)#2 | 77.68(4) |
| I(2)#2-Mo(1)-I(1) | 84.91(3) | Mo(1)#1-I(3)-Mo(1) | 63.30(4) |
| I(3)#1-Mo(1)-Mo(1)#1 | 58.35(2) | Mo(1)#1-I(4)-Mo(1) | 63.31(4) |

Note: Symmetry transformations used to generate equivalent atoms:
　　#1 -x+3/2,-y+1/2,z    #2 -x+3/2,-y+3/2,z

**Table S5:** Anisotropic displacement parameters (A² x 10³)

| | U11 | U22 | U33 | U23 | U13 | U12 |
|---|---|---|---|---|---|---|
| Mo(1) | 31(1) | 39(1) | 30(1) | 0(1) | 0 | 0 |
| I(1) | 43(1) | 37(1) | 32(1) | 0 | 0 | 0 |
| I(2) | 36(1) | 37(1) | 42(1) | 0 | 6(1) | 0 |
| I(3) | 37(1) | 37(1) | 45(1) | 0 | 7(1) | 0 |
| I(4) | 54(1) | 38(1) | 31(1) | 0 | 0 | 0 |

Note: The anisotropic displacement factor exponent takes the form: -2 pi² [ h² a*² U11 + ... + 2 h k a* b* U12 ]

# 4  Exchange coupling constant (J) calculations

To calculate the exchange coupling constant, we considered five structures with different spin configurations (one FM and four different AFM configurations) as shown in Figure S3. In these structures, we have doubled the unit cell along the chain direction (b) to introduce the different AFM spin configurations. The total energy of all these five structures can be derived from the magnetic Hamiltonian as:

$E_{FM} = E_0 + 8S^2J_1 + 8S^2J_2 + 48S^2J_3$
$E_{AFM1} = E_0 - 8S^2J_1 - 8S^2J_2 + 48S^2J_3$
$E_{AFM2} = E_0 - 8S^2J_1 - 8S^2J_2 - 16S^2J_3$
$E_{AFM3} = E_0 - 8S^2J_1 + 8S^2J_2 + 16S^2J_3$
$E_{AFM4} = E_0 + 8S^2J_1 + 8S^2J_2 - 16S^2J_3$

where the exchange constants $J_1$, $J_2$, and $J_3$ are defined in Fig. S3(a), and $E_0$ is the spin-independent term. Utilizing these equations and taking S = 1, we obtain the three exchange coupling constants:

$J_1 + J_2 = (E_{FM} - E_{AFM1}) / 16S^2 = 186.95$ meV
$J_3 = (E_{FM} - E_{AFM4}) / 64S^2 = 0.11873$ meV
$J_1 = (E_{FM} - E_{AFM3} - 32J_3) / 16S^2 = 186.07$ meV
$J_2 = (E_{FM} - E_{AFM1}) / 16S^2 - J_1 = 0.8836$ meV

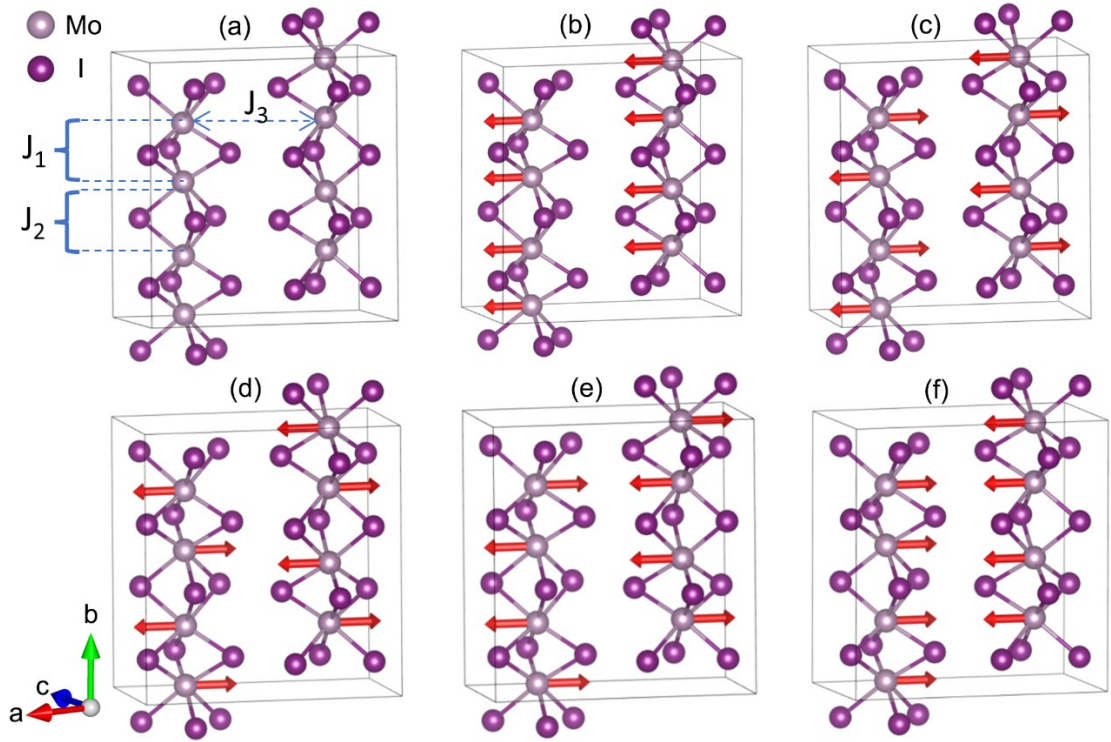

**Figure S3:** (a) MoI₃ supercell (doubled along chain direction) showing the exchange coupling constants. (b)-(e) Shows different spin coupling configurations, (b) FM, (c) AFM1, (d) AFM2, (e) AFM3, (f) AFM4 respectively.

# 5  Magnetic anisotropy energy (MAE) calculation

Here, the MAE is defined as the difference of total energy when the magnetic moments are aligned along the chain direction ($\pm b$) and perpendicular to chain direction ($\pm a$ or $\pm c$). The two values (per formula unit consisting of 4 Mo atoms) are $\mathrm{MAE}_{b,a} = E_{total}(m||\pm b) - E_{total}(m||\pm a) = 0.702$ meV and $\mathrm{MAE}_{b,c} = E_{total}(m||\pm b) - E_{total}(m||\pm c) = 0.647$ meV. This indicates that the moments align perpendicular to the chain in the a-c plane, and within the a-c plane, the magnetic anisotropy is very small. In Eq. (1) of the main text, we have used the average value of MAE divided by the number of Mo atoms in the unit cell, A = 0.67 meV / 4.

**Disclaimer:** Certain commercial equipment, instruments, software or materials are identified in this paper in order to specify the experimental procedure adequately. Such identifications are not intended to imply recommendation or endorsement by National Institute of Standards and Technology (NIST), nor it is intended to imply that the materials or equipment identified are necessarily the best available for the purpose.